\RequirePackage{fixltx2e}
\documentclass[aps,prl,reprint,superscriptaddress,floatfix]{revtex4-1}
\usepackage{graphicx,graphics,amsmath,amssymb,rotate,textcomp,gensymb}
\usepackage{mathrsfs,float}
\usepackage[T1]{fontenc}
\usepackage{dcolumn}
\usepackage{wrapfig}
\usepackage{caption,subcaption}
\usepackage[none]{hyphenat}
\usepackage{hyperref}
\usepackage{times}
\usepackage{natbib}
\usepackage{multirow}

\begin{document}

\title{Raft-like domain formation in model lipid-cholesterol aggregation}

\author{Tanay Paul}
\email{tanaypaul9492@gmail.com}
\author{Jayashree Saha}
\email{jsphy@caluniv.ac.in}
\affiliation{Department of Physics, University of Calcutta,92, A. P. C. Road, Kolkata - 700009, India.}


\begin{abstract}
To understand mechanism of cell-membrane compartmentalization, we studied coarse-grained model systems consisting of lipid and cholesterol molecules. Cholesterol plays crucial role in lateral phase segregation in the bilayered lipid membrane to form cholesterol rich liquid ordered domains, termed as `rafts'. In this NVT Molecular Dynamics simulation study, we have investigated the role of cholesterol 
in raft-like domain formation. The study 
reveals that the strength of cholesterol interaction is an important factor behind the phase segregation to form cholesterol rich domains, whereas, the relative dipolar strength of lipid and cholesterol molecules controls the formation of multi-bilayered stack.
\end{abstract}

\maketitle

The lipid-bilayer of the plasma membrane, which provides basic structure to the eukaryotic cells, is sometimes termed as ``Nature's preferred liquid crystal'' \cite{mouritsenchapter} as it forms lyotropic smectic liquid crystal phase. Extensive experimental studies of the interactions between the components of the bilayered plasma membrane has revealed that it is comprised of laterally segregated and morphologically distinct domains known as `lipid rafts'. The lipid rafts are considered to be relatively small domains, which are depleted of phospholipids with unsaturated acyl chains but enriched in cholesterols and lipids having saturated acyl chains. 
Cholesterol-lipid interactions have long been believed to be an important factor behind the formation of cholesterol rich rafts \cite{silvius}. Cholesterol, which is present in a high concentration in plasma membrane, has crucial effects on the phase behaviour of such phospholipid membrane, as the introduction of cholesterol to a pure phospholipid bilayer changes the membrane properties to intermediate between liquid-disordered (ld) and liquid-ordered (lo) phases \cite{brown2000}. Further works have suggested that a cholesterol dependent lateral phase separation \cite{lingwood} between these two phases can occur in the binary mixtures of phospholipids with cholesterol. Thus, the co-existence of unsaturated phospholipids rich ld domains and cholesterols rich lo domains has been demonstrated as the structure of plasma membranes by extensive model membrane studies \cite{london2002, kane, dietrich, honigmann}.

Cholesterols, when dispersed in aqueous media, forms crystals instead of bilayers, but when mixed with other lipids can adopt complex phase patterns of lateral organization \cite{silvius, yeagle, yeaglebook, mouritsen95} like the one found in plasma membrane. Inspite of significant advances in theoretical and experimental studies of lipid-cholesterol mixtures, the origin of the microscopic organization of lipid-cholesterol bilayers is still not completely known. From the structural point of view, a cholesterol molecule have amphipathic character, just like a phospholipid, due to the presence of polar head group in addition to a hydrophobic chiral tail \cite{yeagle}, but the dipole moment of the polar head group of a cholesterol is much smaller than that of a phospholipid \cite{gopalakrishna}. Cell-biological and biochemical experiments not only have recognized cholesterol as an important constituent of lipid `rafts' in mammalian cell-membranes but also have found that the membrane cholesterol level is a key factor in regulating raft stability and organization \cite{silvius}. The concentration of the cholesterols, present in the lipid bilayer, is an important controlling factor of many other physiological phenomena also. Cholesterol, as a major component of the cell-membrane, can be present in a wide concentration range, $10\%-45\%$, in the cell-membranes. Though the role of the concentration of cholesterol in domain formation was not known properly earlier, more recent studies of the mixture of cholesterols with saturated and unsaturated lipids have revealed that the presence of cholesterol at a typical concentration of $33$ mol$\%$ can create lo domains or rafts \cite{london2002}.

The self-assembly process of amphiphilic molecules like lipids and cholesterols, which involves a number of complex phenomena, is due to a fine competition between different forces of physical origin \cite{cates}. Due to the considerable complexity of plasma membrane and model membranes, computer simulation is an essential technique to study the self-organization and phase-properties of the membrane components \cite{mouritsenchapter}. Some lattice-model simulation studies has been done \cite{nielsen96, nielsen99, banerjee} exploring the phase-equilibria in two-dimensional system of particles considering model interactions between them which imitates different interactions between lipids and cholesterols qualitatively. Effect of the cholesterol concentration on domain formation has been studied varying the percentage of cholesterol molecules in a model system of multi-component lipid bilayer using random lattice-model simulation allowing both translational and internal degrees of freedom \cite{banerjee}. Various computer simulation investigations \cite{saiz, brannigan, cooke, arnarez, marrink, farago, sodt, allen, whitehead, ayton, sun} on large-scale lipid membrane properties has also been done using coarse-grained generic models instead of all-atom models as the latter takes huge computation time. 

In this Molecular Dynamics (MD) simulation study of lipid-cholesterol mixture, coarse-grained modelling of lipid and cholesterol molecules have been done to study the phase-equilibria and the dependence of `raft' formation on the presence of cholesterol. As both the lipid and chiral molecules have a polar head-group part as well as long apolar tail, they have been modelled as ellipsoidal particles which are interacting via van-der-Waal type interaction, known as Gay-Berne (GB) potential \cite{GB}. In the field of liquid crystal, this GB force field has been widely used to investigate various phases \cite{allenreview}. In some previous MD simulation study of lipid bilayers \cite{whitehead, ayton, sun} also, GB ellipsoids have been used to model large length-scale properties. In our model, the polar interaction between two lipid molecules or two cholesterol molecules or one lipid and one cholesterol molecule has been taken as simply dipole-dipole interaction. More over, as cholesterol molecules are chiral in nature, two cholesterol molecules have been considered as interacting via a model chiral interaction \cite{paul19}. Thus a lipid molecule has been modelled as an achiral ellipsoidal molecule of length to breadth ratio $3:1$ and a cholesterol molecule has been modelled as a polar chiral molecule of ellipsoidal shape and of same size, with a point dipole embedded at $0.5\sigma_0$ distance from one end of both type of molecules, $\sigma_0$ being the breadth of the molecules. Direction of the dipole is fixed at $90\degree$ with respect to the molecular long axis, as the preferential direction of the dipolar head-groups of the lipid particles constituting bilayered membrane is to lie on the plane of the bilayer \cite{paul17}.

Thus two such lipid molecules and one lipid with another cholesterol molecule interact between them via the pair potential given by,
\begin{eqnarray}
U_{1}(\vec{r}_{ij},\hat{u}_{i},\hat{u}_{j})&=& 4\epsilon(\hat{r}_{ij}, \hat{u}_{i}, \hat{u}_{j})(\rho_{ij}^{-12}-\rho_{ij}^{-6}) \nonumber \\
& & + \frac{1}{r^3_{d}}[\vec{\mu}_{d_{i}}\cdot\vec{\mu}_{d_{j}}-\frac{3}{r^{2}_{d}}(\vec{\mu}_{d_{i}}\cdot\vec{r}_{d})(\vec{\mu}_{d_{j}}\cdot\vec{r}_{d})] \nonumber \\
&=& U_{GB}(\vec{r}_{ij},\hat{u}_{i},\hat{u}_{j}) + U_{dd}(\vec{r}_{d},\hat{u}_{d_{i}},\hat{u}_{d_{j}}) \label{eq:U1}
\end{eqnarray}
whereas, the pair interaction between such two cholesterol molecules is taken as,
\begin{eqnarray}
U_{2}(\vec{r}_{ij},\hat{u}_{i},\hat{u}_{j})&=& 4\epsilon(\hat{r}_{ij}, \hat{u}_{i}, \hat{u}_{j})(\rho_{ij}^{-12}-\rho_{ij}^{-6}) \nonumber \\
& & -c\ 4\epsilon(\hat{r}_{ij}, \hat{u}_{i}, \hat{u}_{j})\rho_{ij}^{-7}\{(\hat{u}_{i}\times\hat{u}_{j})\cdot\hat{r}_{ij}\}(\hat{u}_{i}\cdot\hat{u}_{j}) \nonumber \\
& & +\frac{1}{r^3_{d}}[\vec{\mu}_{d_{i}}\cdot\vec{\mu}_{d_{j}}-\frac{3}{r^{2}_{d}}(\vec{\mu}_{d_{i}}\cdot\vec{r}_{d})(\vec{\mu}_{d_{j}}\cdot\vec{r}_{d})] \nonumber \\
&=& U_{GB}(\vec{r}_{ij},\hat{u}_{i},\hat{u}_{j})+\ U_{c}(\vec{r}_{ij},\hat{u}_{i},\hat{u}_{j}) \nonumber \\
& & +U_{dd}(\vec{r}_{d},\hat{u}_{d_{i}},\hat{u}_{d_{j}})  \label{eq:U2}
\end{eqnarray}
Here, $U_{GB}$, the well known GB potential \cite{GB, luckhurst}, is the molecular position and orientation dependent achiral potential of van-der-Waal type. $U_{c}$ is the chiral interaction potential \cite{memmer} which induces a twist angle between two side-by-side ellipsoidal molecules and energetically favours parallel arrangement for two end-to-end molecules, thus gives rise to chiral phases. The sign of the chirality strength parameter $c$ determines the handedness. In the expressions of both $U_{GB}$ and $U_{c}$, the term $\rho_{ij}$ is given by, $\rho_{ij} = [r_{ij} - \sigma(\hat{r}_{ij}, \hat{u}_{i}, \hat{u}_{j}) + \sigma_{0}] / \sigma_{0}$. Here, $\sigma(\hat{r}_{ij}, \hat{u}_{i}, \hat{u}_{j})$ is the orientation dependent separation term and $\epsilon(\hat{r}_{ij}, \hat{u}_{i}, \hat{u}_{j})$ is the orientation dependent well depth between two molecules $i$ and $j$. The vector $\vec{r}_{ij}=r_{ij}\hat{r}_{ij}$ is the separation between $i$-th and $j$-th molecules, $\hat{u}_{i}$ and $\hat{u}_{j}$ being the unit vectors along the long axis of the respective molecules. The values of the GB parameters used in our study are $\kappa=3.0$, $\kappa'=1/5$, $\mu=1$, $\nu=2$ for both types of molecules. Lastly, $U_{dd}$ is simply a dipole-dipole interaction potential separated by distance $\vec{r}_{d}$. The dipole moment vector of the point dipole embedded on $i$-th molecule is given by, $\vec{\mu}_{d_{i}}\equiv \mu^*\hat{u}_{d_{i}}$, where $\mu^*= (\mu^2/\varepsilon_{s}\sigma_{0}^3)^{1/2}$ is the scaled dipole moment strength, the value of which has been set fixed to $1.4$ for a lipid molecule while different set of values of $\mu^*$ has been considered for the model molecule of cholesterol. To consider the long-range nature of the dipolar interaction conventional reaction-field technique has been used within a sphere of cut-off radius $r_{RF}=0.5/\sigma_0$ and taking the dielectric constant of the continuum outside it as $\epsilon_{RF}=1.5$ \cite{berardi99}. To reduce computation time, explicit water interaction has not been considered in this simulation study.

In this NVT-MD simulation study, the scaled density $\rho^*$ has been set fixed to an optimum value of $0.30$. For each system, an well equilibrated isotropic phase consisting both types of molecules has been used as the initial configuration. The scaled temperature $T^*(=k_{B}T/\epsilon_{s}\text{, }k_{B}\text{ being the Boltzmann Constant})$ has been decreased gradually to get the equilibrium stable phases at lower temperatures and at a particular temperature stage, the simulation run has been started from a higher temperature equilibrium phase. At each temperature stage, to realize the equilibrium phase, simulation run of nearly $10^6$ MD-steps has been performed and it has been ensured that the rms fluctuation of the average energy of the system remains within $2\%$ about a mean value at equlibrium. Production run of another $5\times 10^4$ steps has been performed to calculate required averages. The concentration of cholesterol, i.e., chiral polar molecules, has been taken as $30\%$ of the total number of molecules ($N$) and in the initial isotropic phase both types of molecules have been taken as distributed randomly over the whole simulation box. Simulation results for $N=864$ \& $1372$ molecules have been compared which shows qualitatively same behaviours.

\begin{figure}[h]
\begin{center}
 \begin{subfigure}{0.24\textwidth}
  \includegraphics[width=\textwidth]{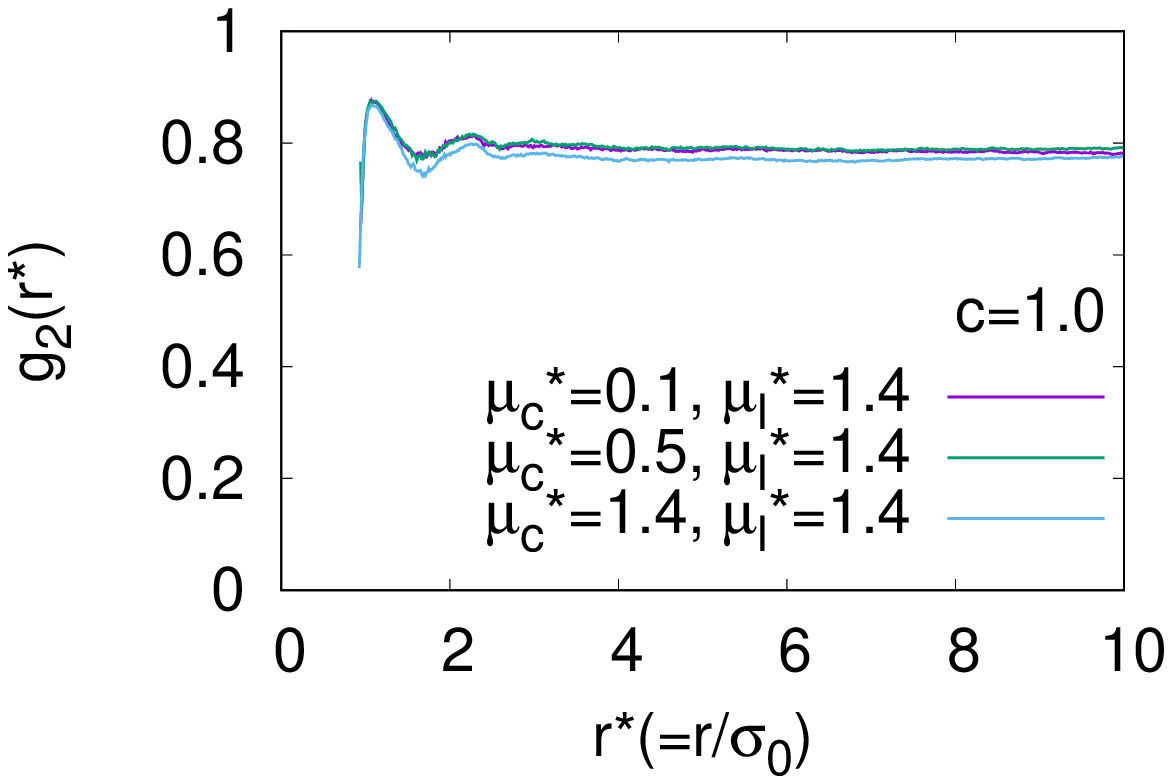}
  \caption{\label{fig:g2ofr_ch1.0}}
 \end{subfigure}\hfill
 \begin{subfigure}{0.24\textwidth}
  \includegraphics[width=\textwidth]{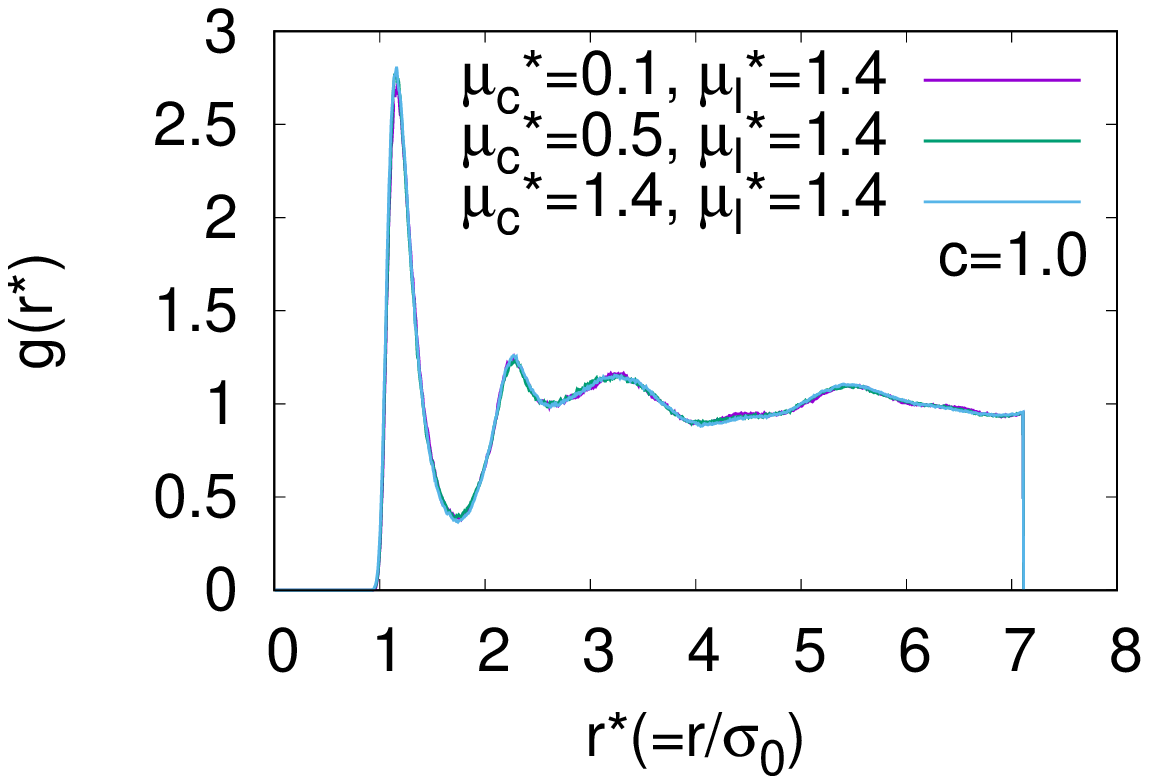}
  \caption{\label{fig:gofr_ch1.0}}
 \end{subfigure}\\
 \begin{subfigure}{0.24\textwidth}
  \includegraphics[width=\textwidth]{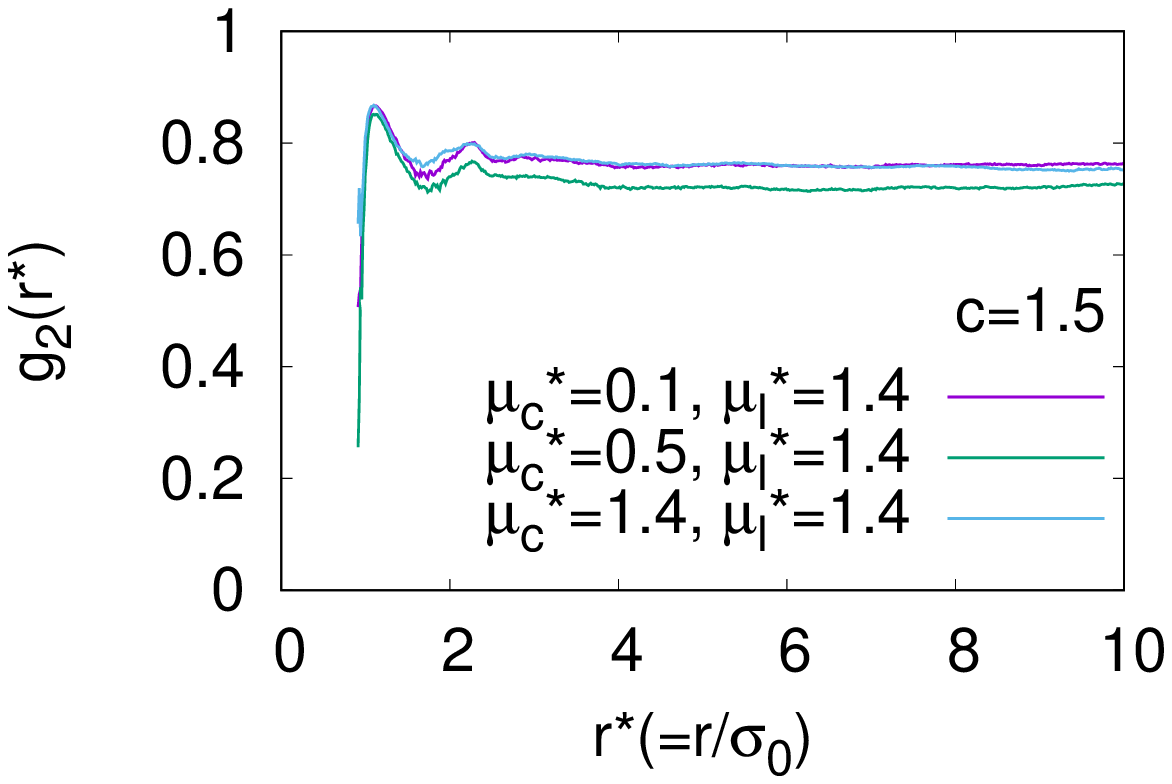}
  \caption{\label{fig:g2ofr_ch1.5}}
 \end{subfigure}\hfill
 \begin{subfigure}{0.24\textwidth}
  \includegraphics[width=\textwidth]{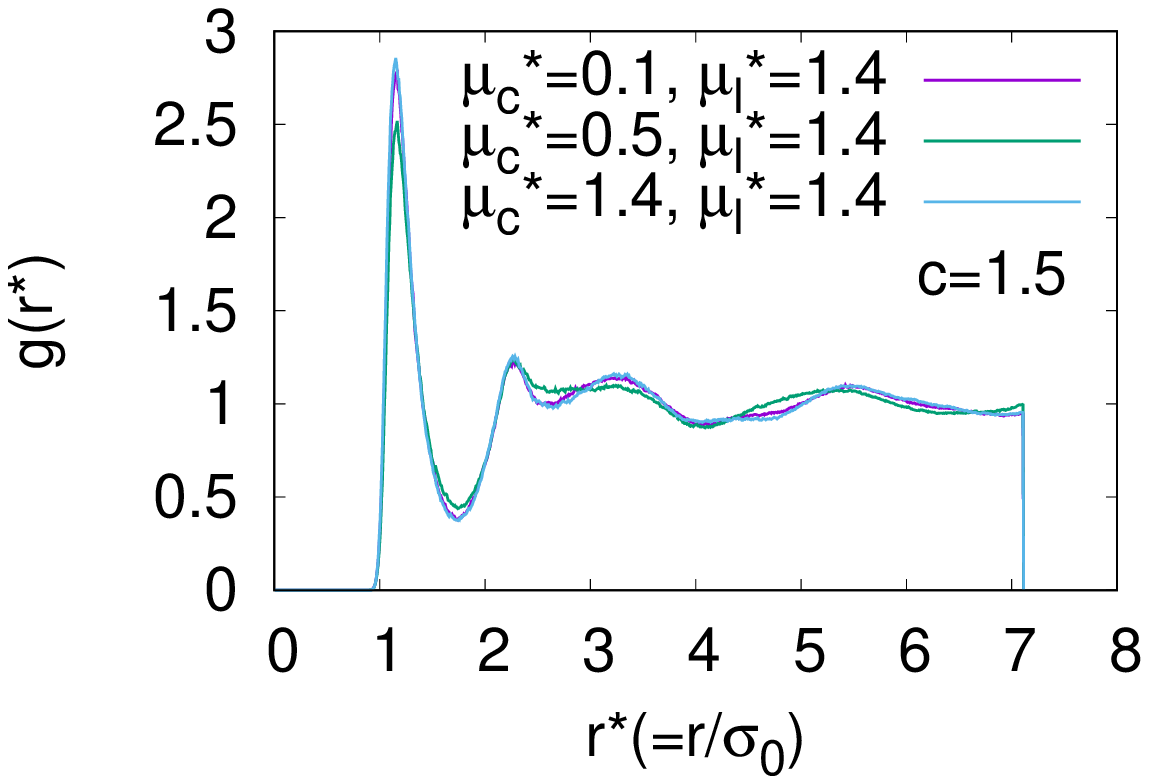}
  \caption{\label{fig:gofr_ch1.5}}
 \end{subfigure}
 \caption{Plots of (a) $g_{2}(r^*)$ and (b) $g(r^*)$ for $c=1.0$; (c) $g_{2}(r^*)$ and (d) $g(r^*)$ for $c=1.5$.\label{fig:g2ofr_gofr}}
\end{center}
\end{figure}

\begin{figure}[h]
\begin{center}
 \begin{subfigure}{0.21\textwidth}
  \includegraphics[width=\textwidth]{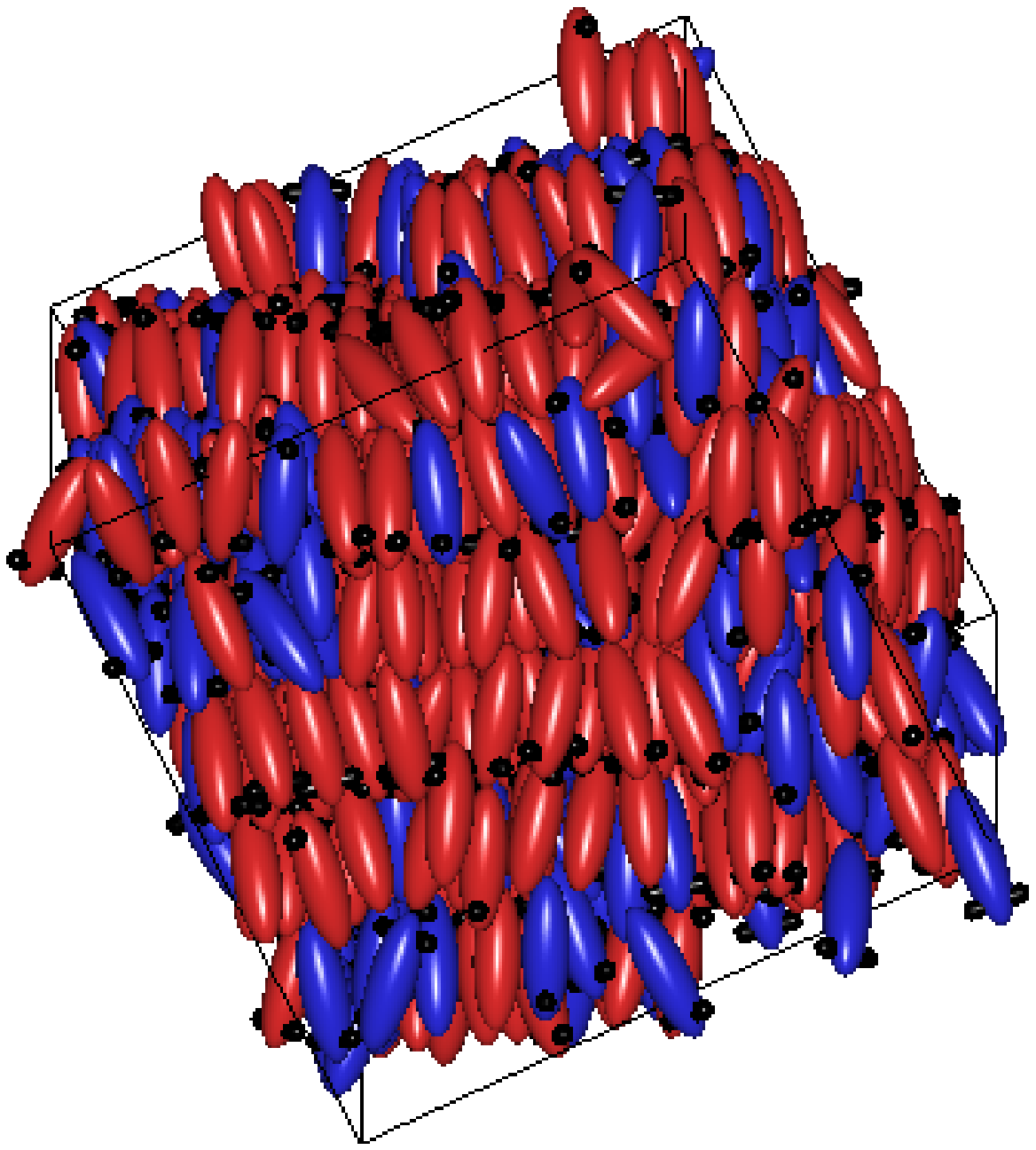}
  \caption{\label{qmga_ch1.0_1.4_0.1}}
 \end{subfigure}
 \begin{subfigure}{0.23\textwidth}
  \includegraphics[width=\textwidth]{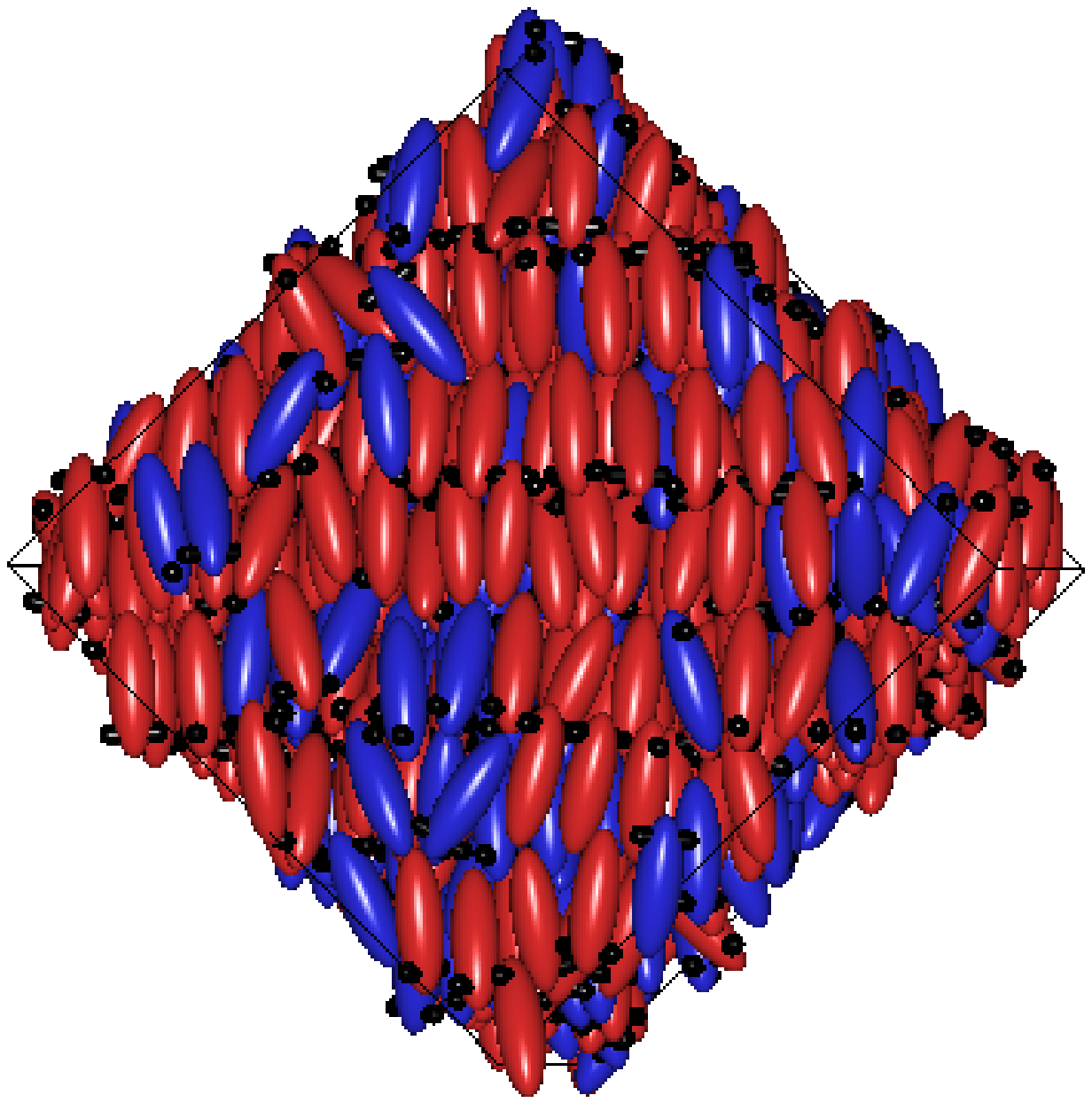}
  \caption{\label{qmga_ch1.0_1.4_0.5}}
 \end{subfigure}\\
 \begin{subfigure}{0.23\textwidth}
  \includegraphics[width=\textwidth]{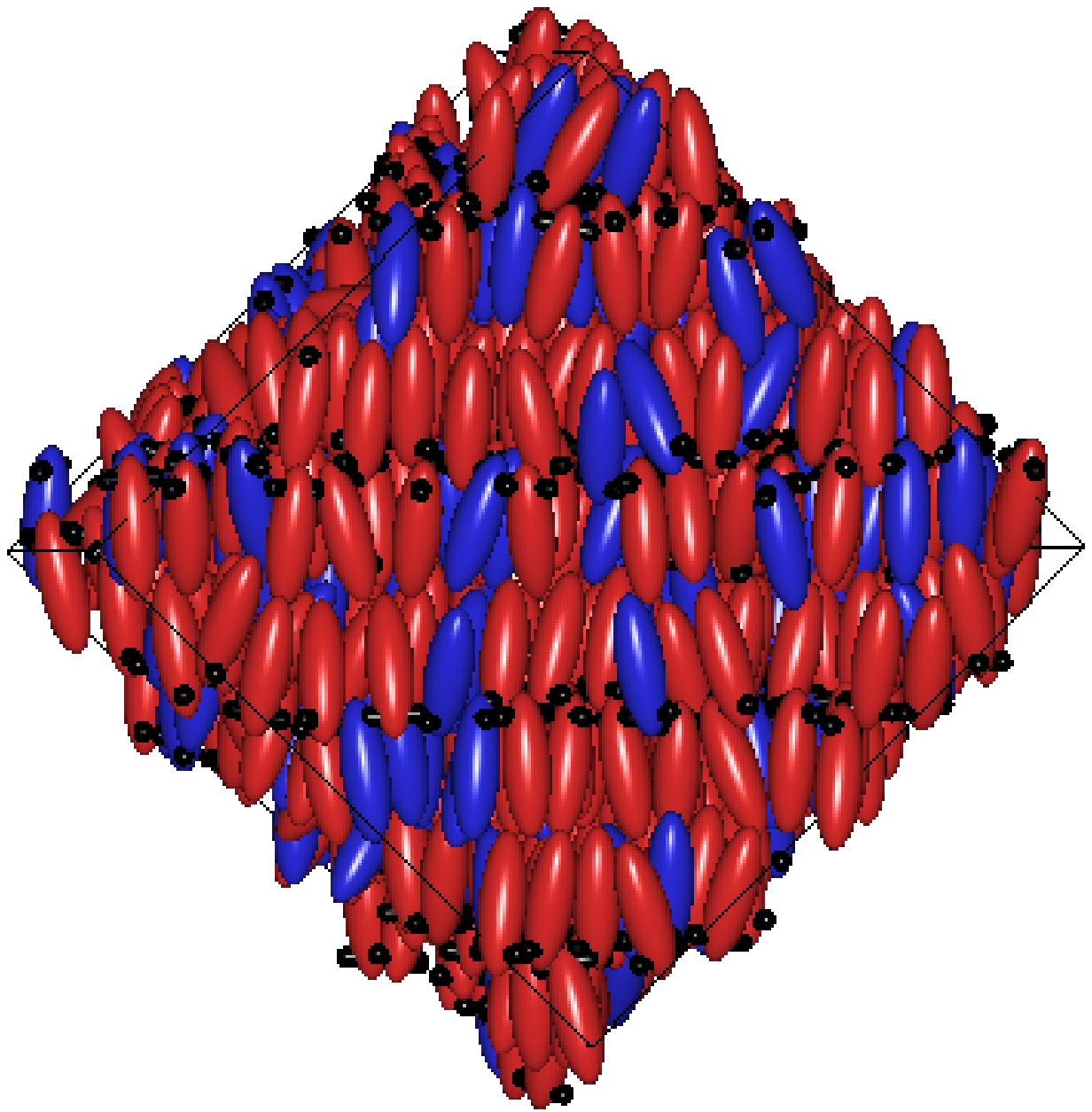}
  \caption{\label{qmga_ch1.0_1.4_1.4}}
 \end{subfigure}
 \begin{subfigure}{0.23\textwidth}
  \includegraphics[width=\textwidth]{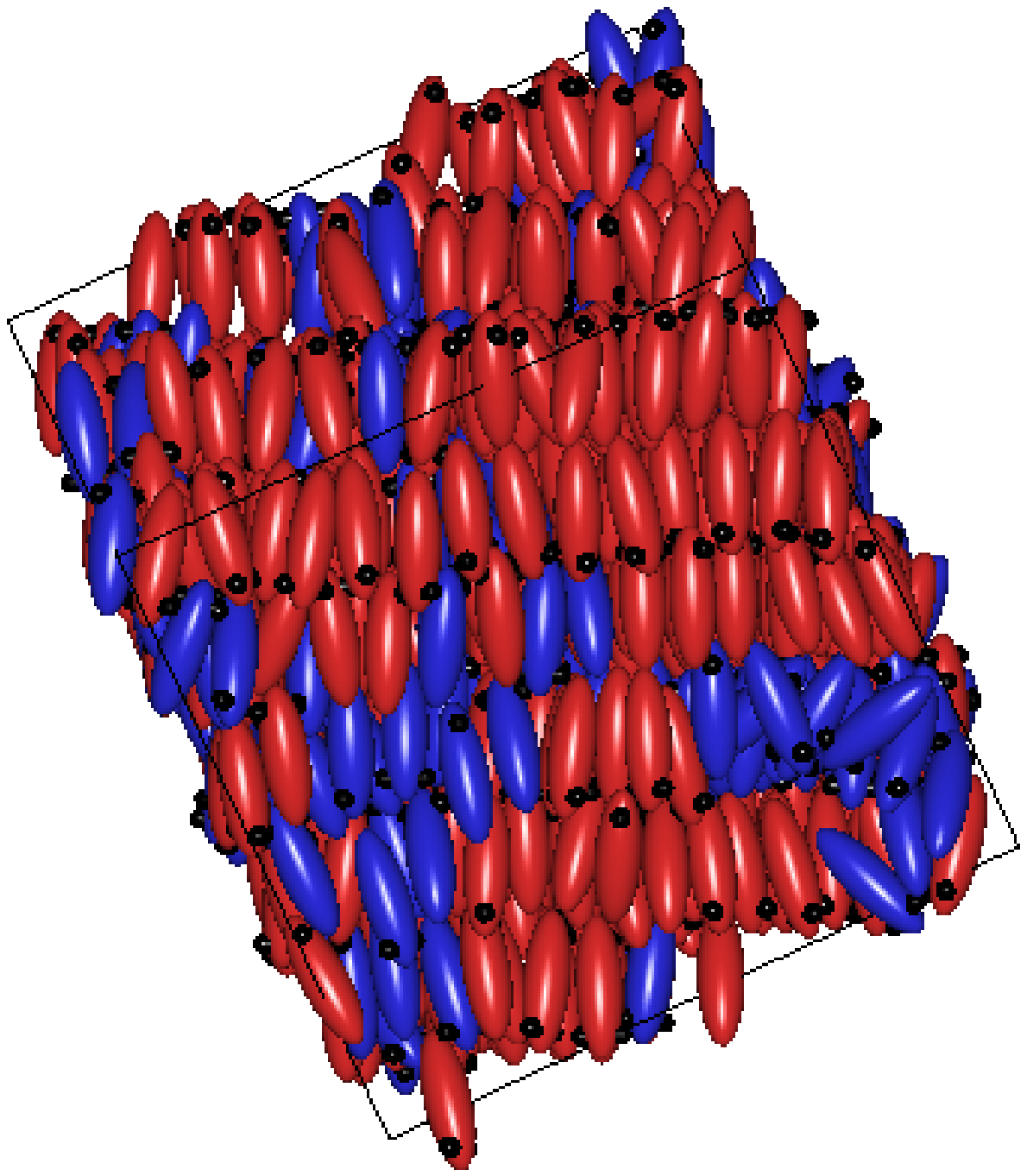}
  \caption{\label{qmga_ch1.5_1.4_0.1}}
 \end{subfigure}\\
 \begin{subfigure}{0.23\textwidth}
  \includegraphics[width=\textwidth]{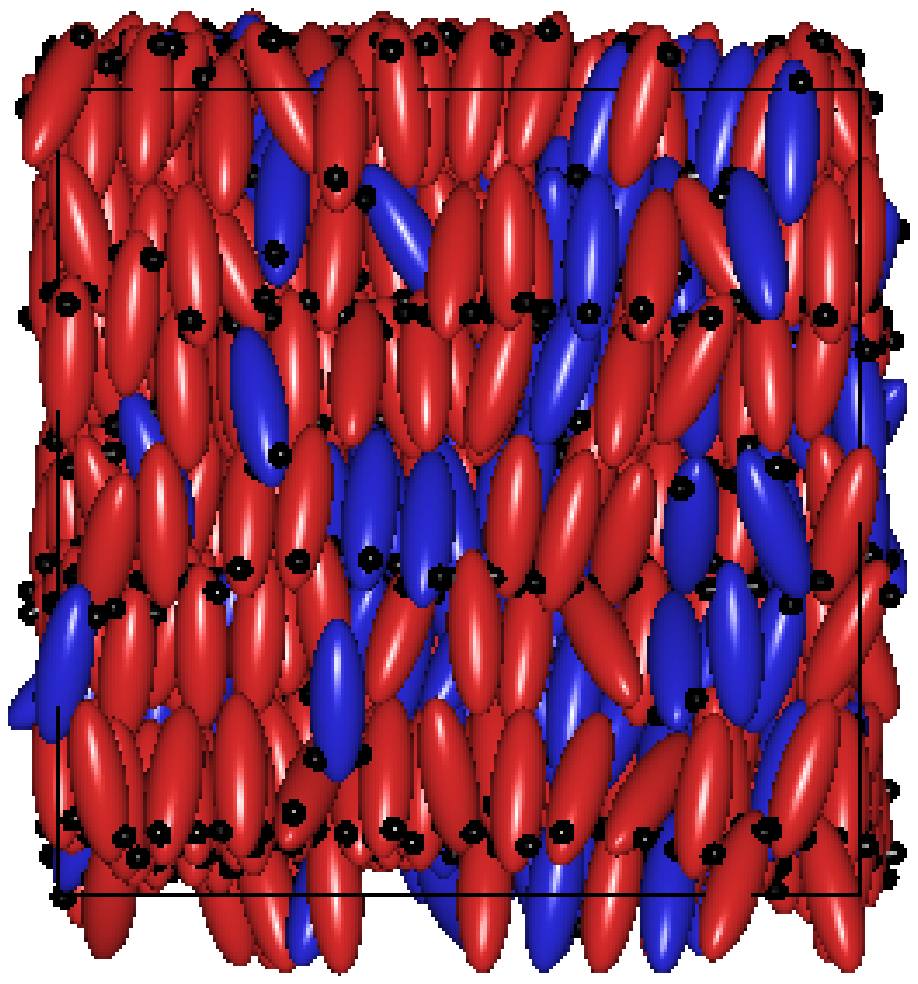}
  \caption{\label{qmga_ch1.5_1.4_0.5}}
 \end{subfigure}
 \begin{subfigure}{0.23\textwidth}
  \includegraphics[width=\textwidth]{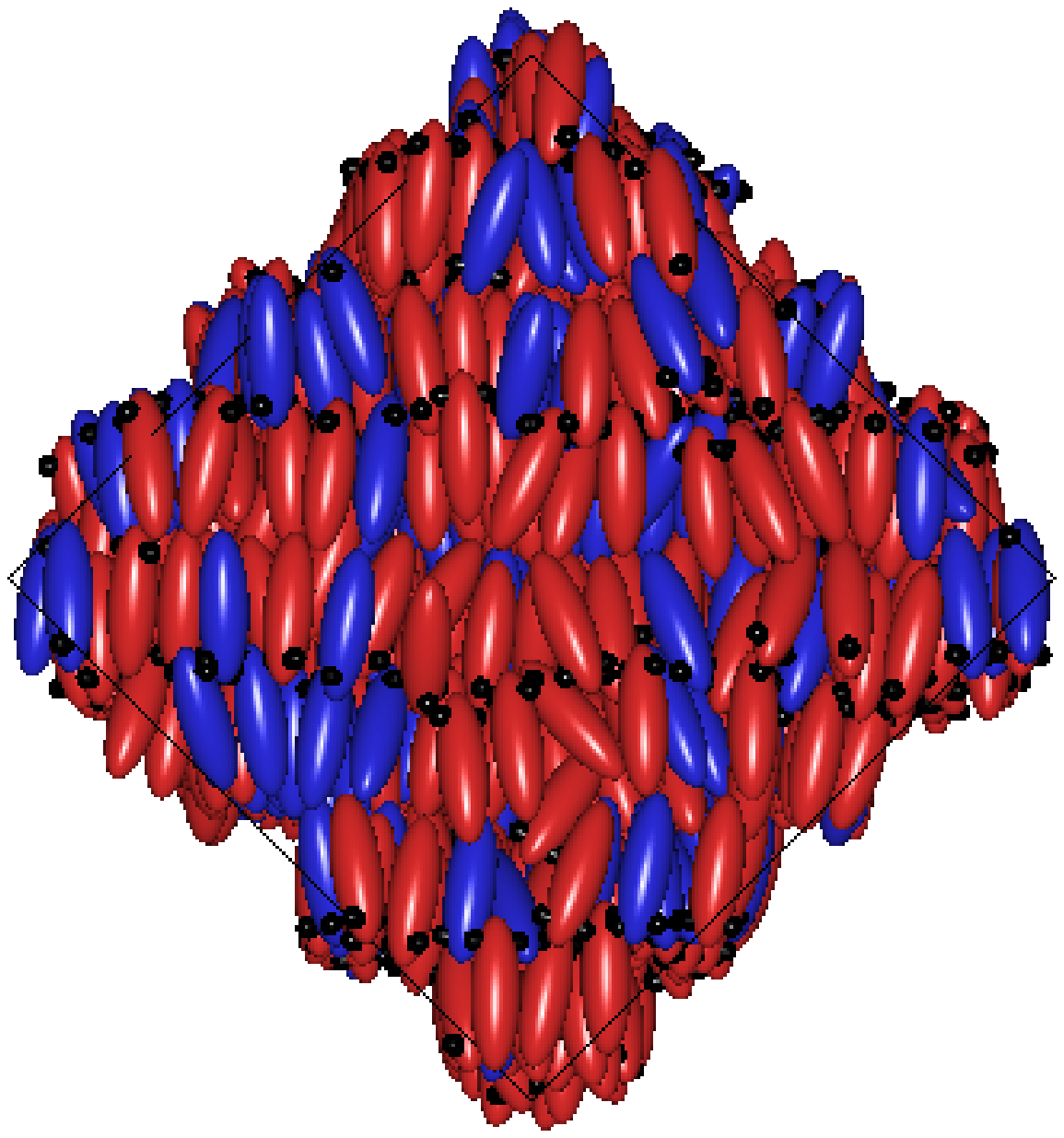}
  \caption{\label{qmga_ch1.5_1.4_1.4}}
 \end{subfigure}
 \caption{Snapshots of the configurations for $c=1.0$, $\mu^*_l=1.4$, (a) $\mu^*_c=0.1$, (b) $\mu^*_c=0.5$, (c) $\mu^*_c=1.4$ and for $c=1.5$, $\mu^*_l=1.4$, (d) $\mu^*_c=0.1$, (e) $\mu^*_c=0.5$, (f) $\mu^*_c=1.4$. Polar achiral molecules are shown in red and polar chiral molecules in blue. Positions of the dipoles are shown in black dots.}
\end{center}
\end{figure}

The strength of the chiral interaction, i.e. the value of the chirality strength parameter $c$, is an important factor which can control the phase properties of the assembly of polar chiral molecules \cite{paul19}. In this study of the mixture of polar chiral and achiral molecules, the effect of the chirality strength parameter $c$ has been checked by considering two different values of $c$ and performing separate simulation study. The results have been checked for two separate values of the scaled dipole moment for chiral molecules ($\mu^*_c$, say), $0.1$ and $0.5$, whereas that of the achiral polar molecules ($\mu^*_l$, say), representing lipid molecules, has been set to a relatively larger value $1.4$. When the value of $c$ is set as $1.0$, for both the systems having $\mu^*_c=0.1$ and $0.5$, nematic ordering has been formed in the system as the scaled temperature $T^*$ has been decreased gradually starting from an isotropic phase. The distribution of the orientational order, $g_2(r^*)=\frac{3}{2}\langle \sum_{i} \sum_{j\neq i}(\hat{u}_{i}(r^*)\cdot \hat{u}_{j}(r^*))\rangle -\frac{1}{2}$ ($r^*$ is the scaled intermolecular separation), has been checked (figure: \ref{fig:g2ofr_ch1.0}) which shows strong orientational correlation over the whole system. The plots of radial distribution function or pair distribution function, $g(r^*)=V\langle \sum_i \sum_{j\neq i}\delta(r^*-r^*_{ij})\rangle /N^2$ \cite{allenbook}, show short range positional order but less order in long range in this phase (figure: \ref{fig:gofr_ch1.0}). On further decrease in the temperature, smectic layers has been formed from nematic phase. But, in this case both types of molecules are mixed in such a way that there exist tiny domains of polar chiral molecules inside the matrix of polar non-chiral molecules. Such domains, though weakly formed, can be found observing the structures of the 
stable phases obtained in this case for different $\mu^*$ values. Snapshots of these phases obtained with $\mu^*_c=0.1$ and $0.5$ has been presented in figures \ref{qmga_ch1.0_1.4_0.1} and \ref{qmga_ch1.0_1.4_0.5} respectively.

\begin{figure}[h!]
\begin{center}
 \begin{subfigure}{0.48\textwidth}
  \begin{subfigure}{0.48\textwidth}
  \includegraphics[width=\textwidth]{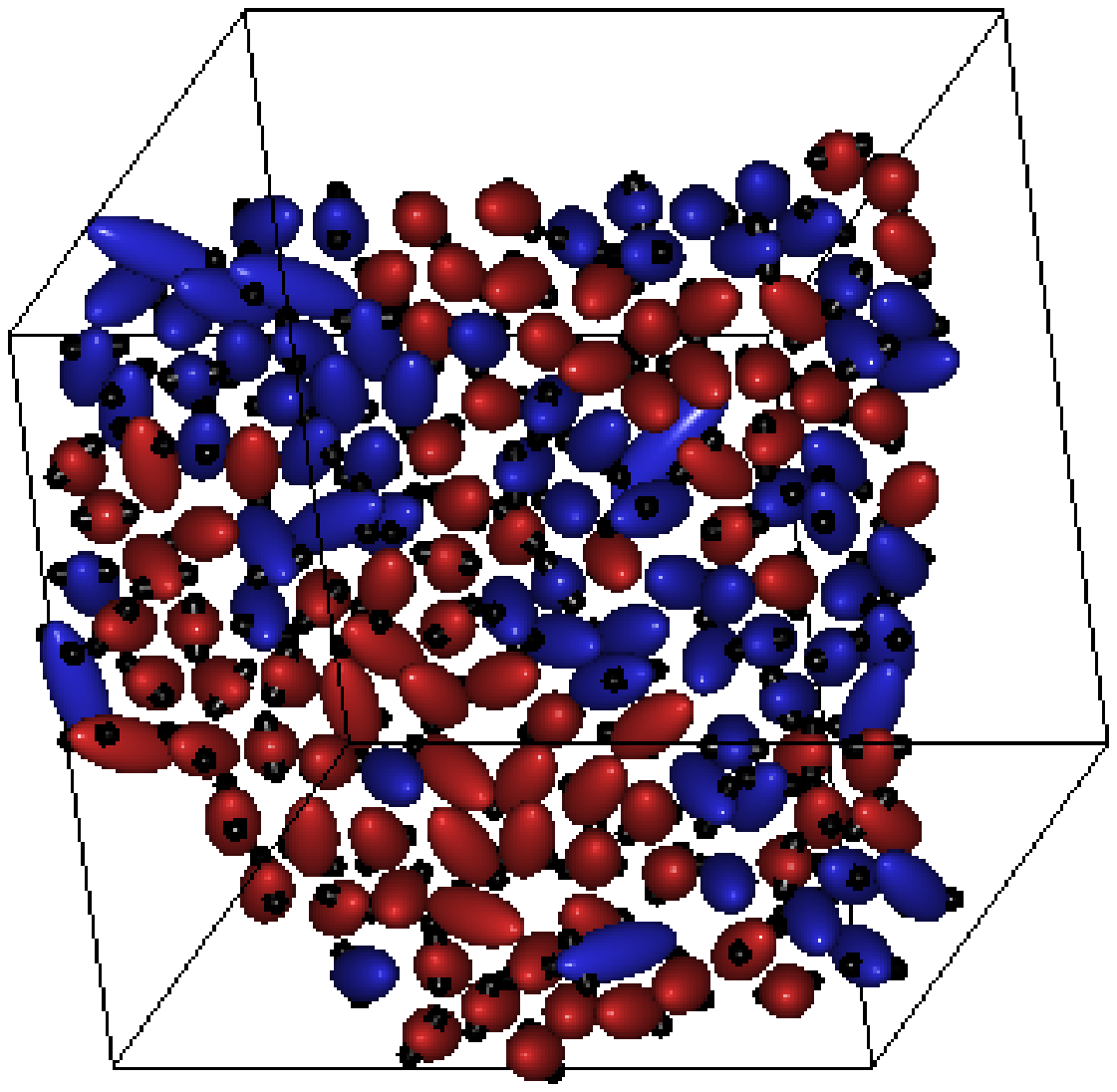}
  \end{subfigure}
  \begin{subfigure}{0.48\textwidth}
  \includegraphics[width=\textwidth]{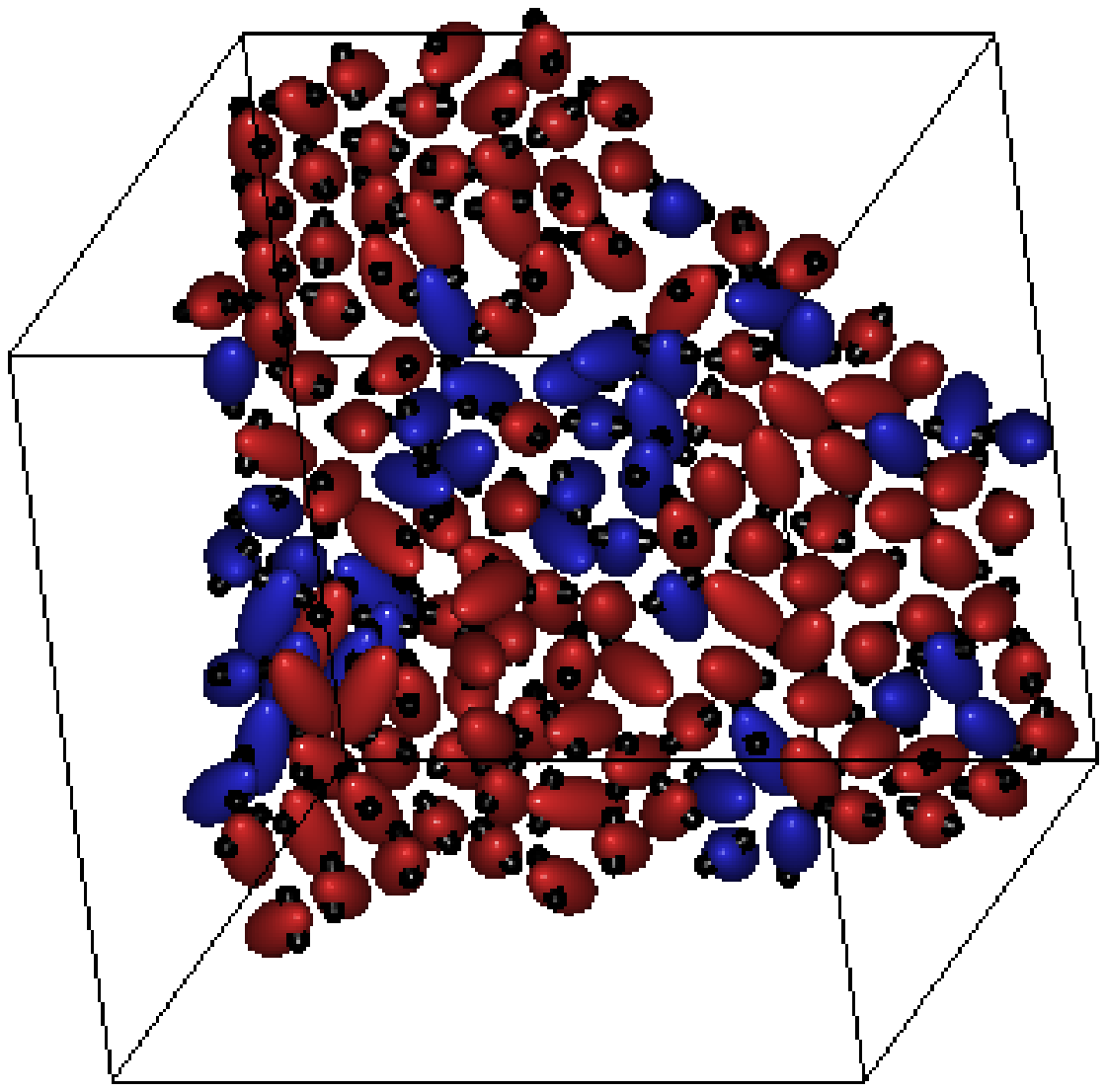}
  \end{subfigure}\\
  \begin{subfigure}{0.48\textwidth}
  \includegraphics[width=\textwidth]{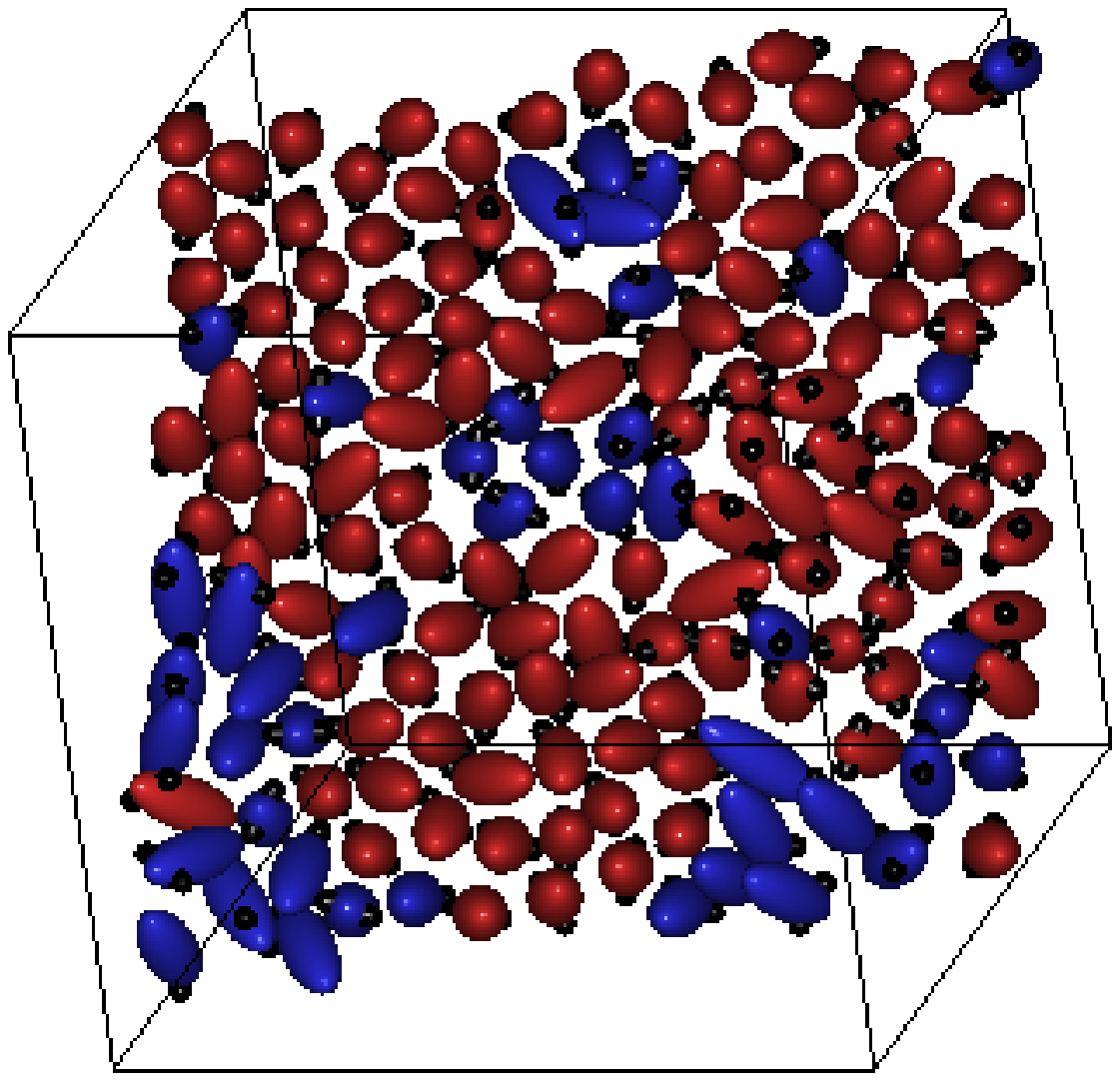}
  \end{subfigure}
  \begin{subfigure}{0.48\textwidth}
  \includegraphics[width=\textwidth]{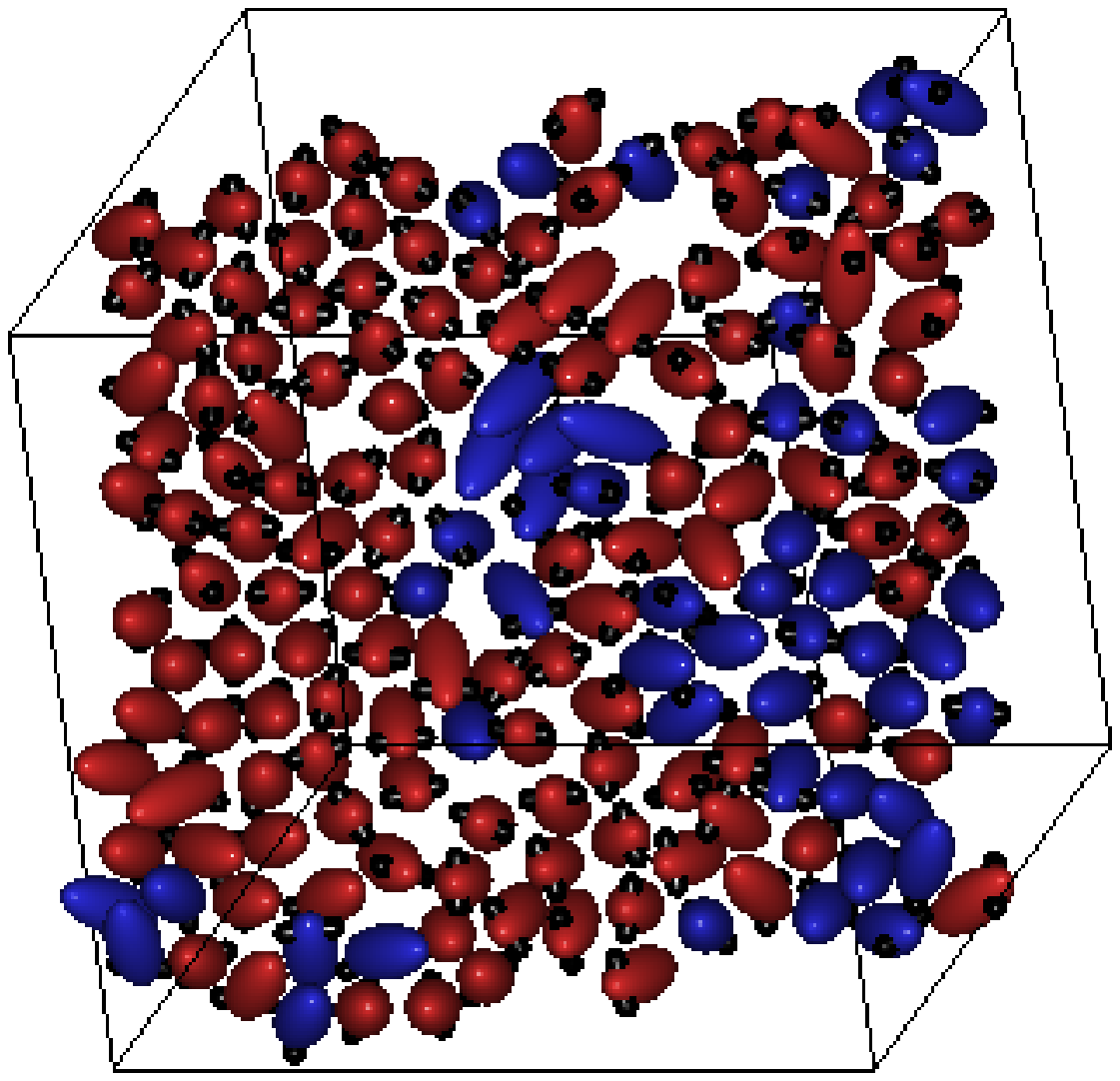}
  \end{subfigure}
  \caption{$c=1.5$, $\mu^*_c=0.1$, $\mu^*_l=1.4$.\label{fig:layer_c1.5_1.4_0.1}}
 \end{subfigure}\\
 \begin{subfigure}{0.48\textwidth}
  \begin{subfigure}{0.44\textwidth}
  \includegraphics[width=\textwidth]{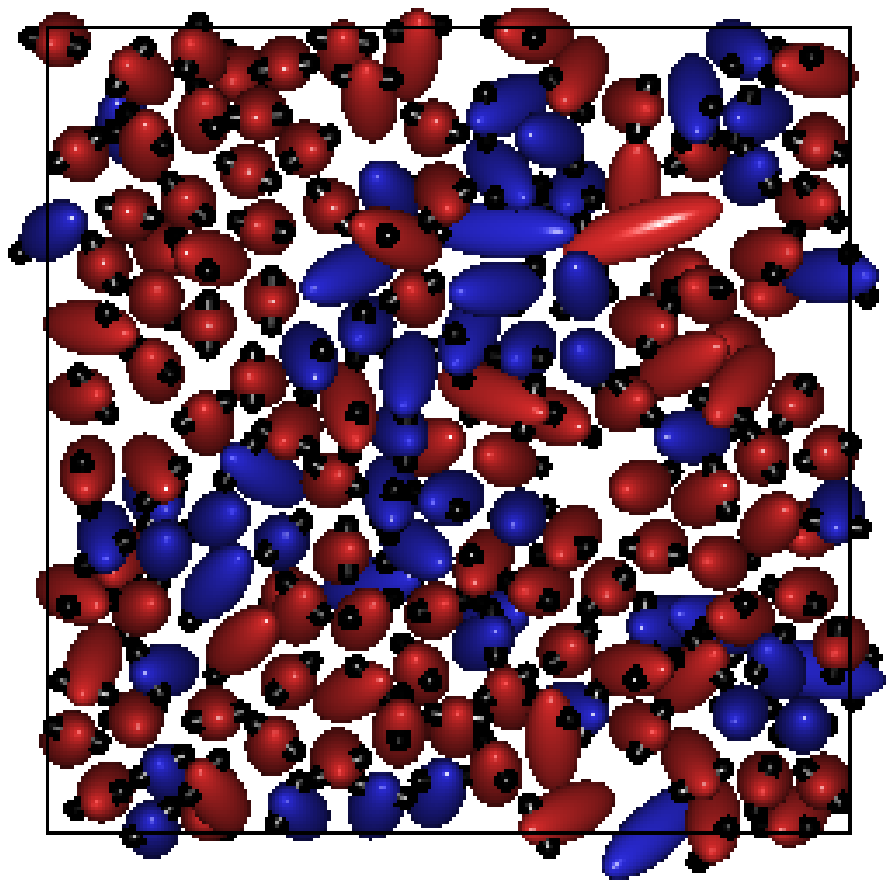}
  \end{subfigure}
  \begin{subfigure}{0.44\textwidth}
  \includegraphics[width=\textwidth]{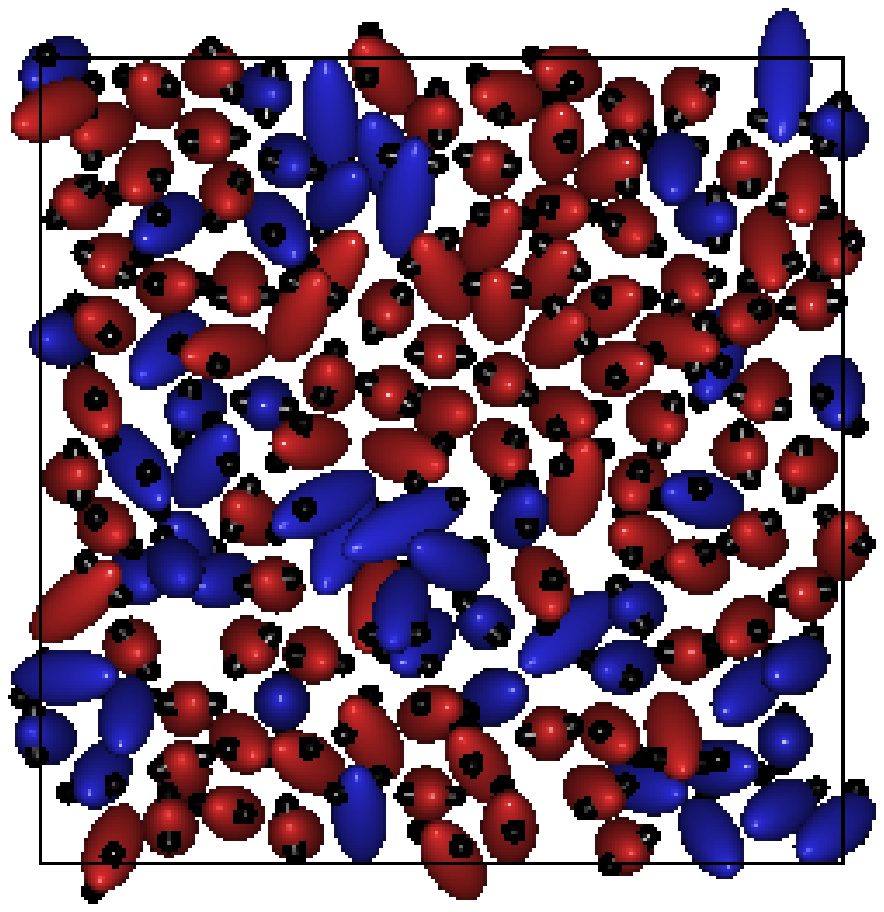}
  \end{subfigure}\\
  \begin{subfigure}{0.44\textwidth}
  \includegraphics[width=\textwidth]{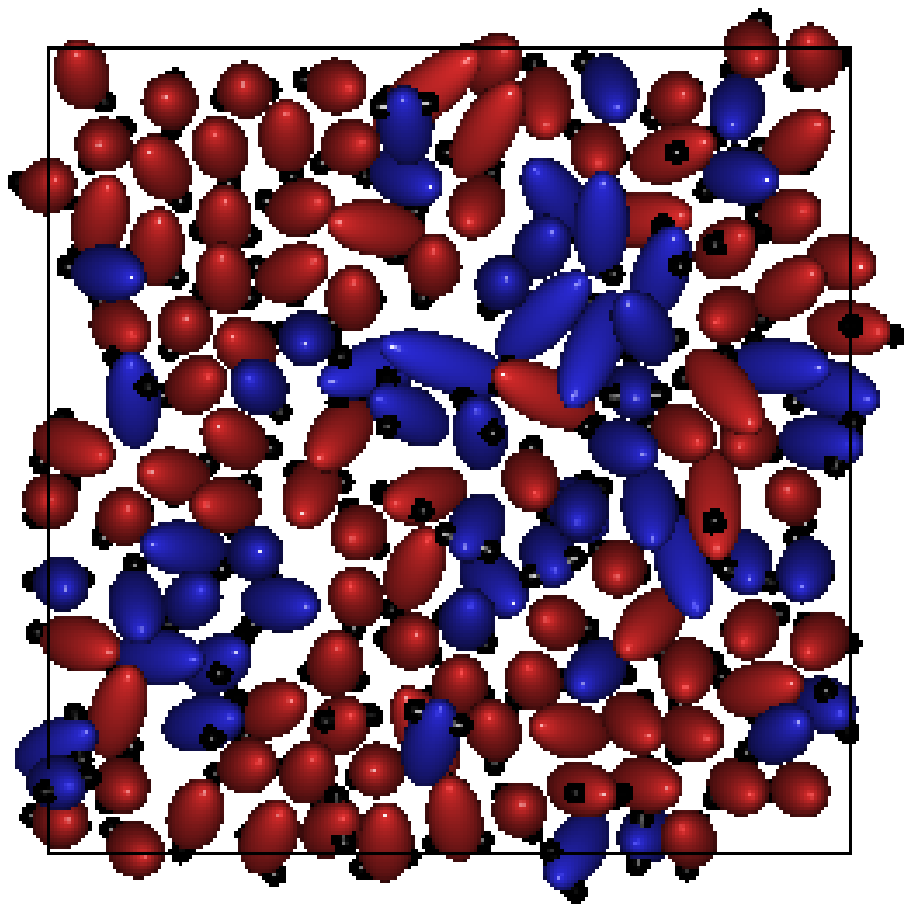}
  \end{subfigure}
  \begin{subfigure}{0.44\textwidth}
  \includegraphics[width=\textwidth]{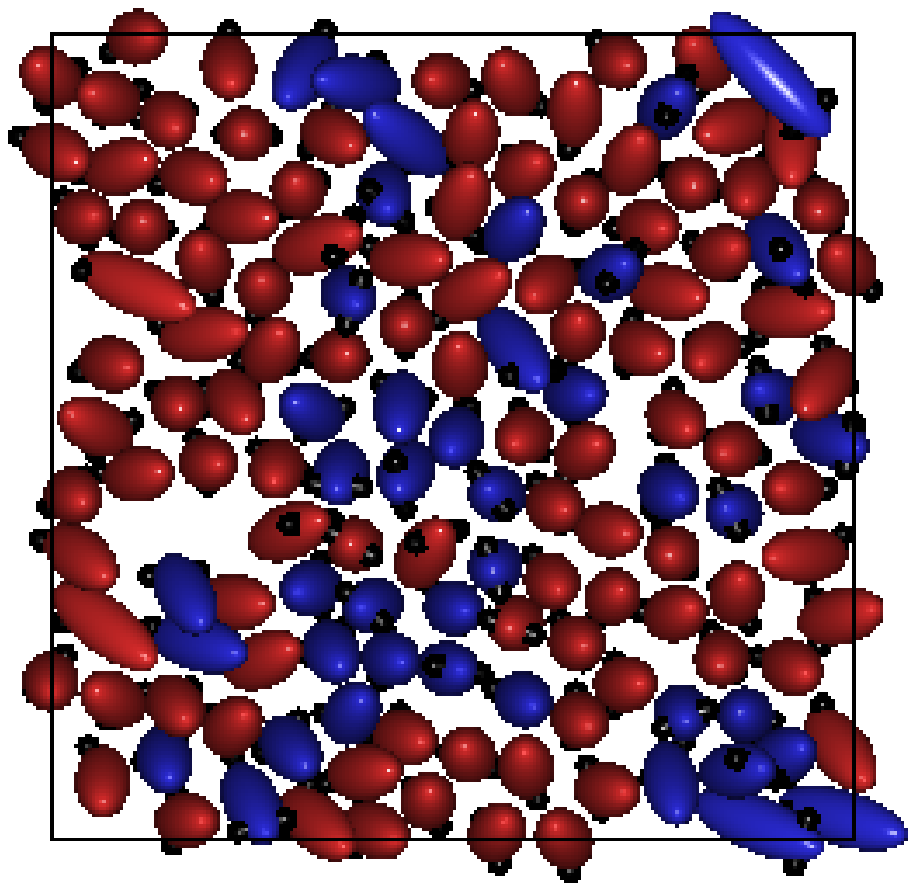}
  \end{subfigure}
  \caption{$c=1.5$, $\mu^*_c=0.5$, $\mu^*_l=1.4$.\label{fig:layer_c1.5_1.4_0.5}}
 \end{subfigure}
 \caption{Snapshots of the configurations of some of the smectic layers obtained with $c=1.5$, $\mu^*_l=1.4$, (a) $\mu^*_c=0.1$, and (b) $\mu^*_c=0.5$, for $N=864$. View from the top of the layers i.e. along the direction of the director axis has been presented to show the particle positions only. Particles in red represent polar achiral molecules and particles in blue are polar chiral molecules.\label{fig:layers_1}}
\end{center}
\end{figure}

When the value of $c$ has been set to $1.5$, with relative concentration of $30\%$ chiral polar molecules, similar nematic ordering has been found to occur in the system by decreasing the temperature from an isotropic phase keeping the density of the system fixed. Plots of the orientational distribution function $g_2(r^*)$ in this case too, show strong long distance orientational order (figure: \ref{fig:g2ofr_ch1.5}) and the plots of pair distribution function $g(r^*)$ (figure: \ref{fig:gofr_ch1.5}) show the similar behaviour qualitatively as in the case with $c=1.0$. Interestingly, in this case it has been found that, the chiral polar molecules have been aggregated together to form small domains of chiral molecules as the system has been allowed to evolve further keeping temperature fixed. Similar phases have been formed for both the values of $\mu^*_c=0.1$ and $0.5$. As the temperature has been decreased further, starting from such a phase, smectic layers have been formed. Snapshots of the configurations of some of the smectic layers of considerable size, obtained in both the cases have been presented in figure \ref{fig:layers_1}, where the small domains of polar chiral molecules can clearly be identified. Comparison of the results obtained with $c=1.0$ and $c=1.5$ shows that the formation of domains enriched with polar chiral molecules is relatively stronger for $c=1.5$. 

\begin{figure}[h]
\begin{center}
 \begin{subfigure}{0.48\textwidth}
 \includegraphics[width=\textwidth, height=0.35\textwidth]{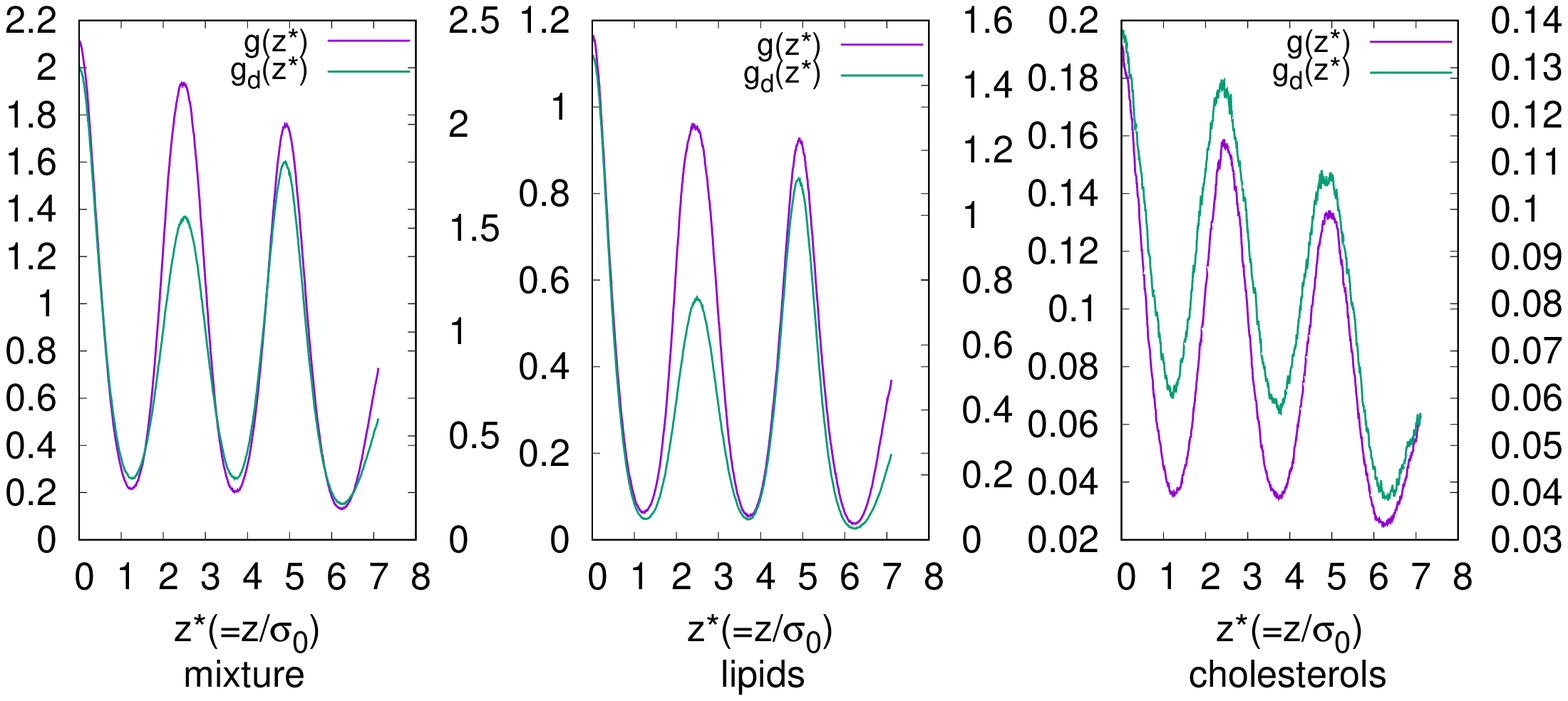}
 \caption{$c=1.5$, $\mu^*_c=0.1$, $\mu^*_l=1.4$.\label{gofz_c1.5_1.4_0.1}}
 \end{subfigure}\\
 \begin{subfigure}{0.48\textwidth}
 \includegraphics[width=\textwidth, height=0.35\textwidth]{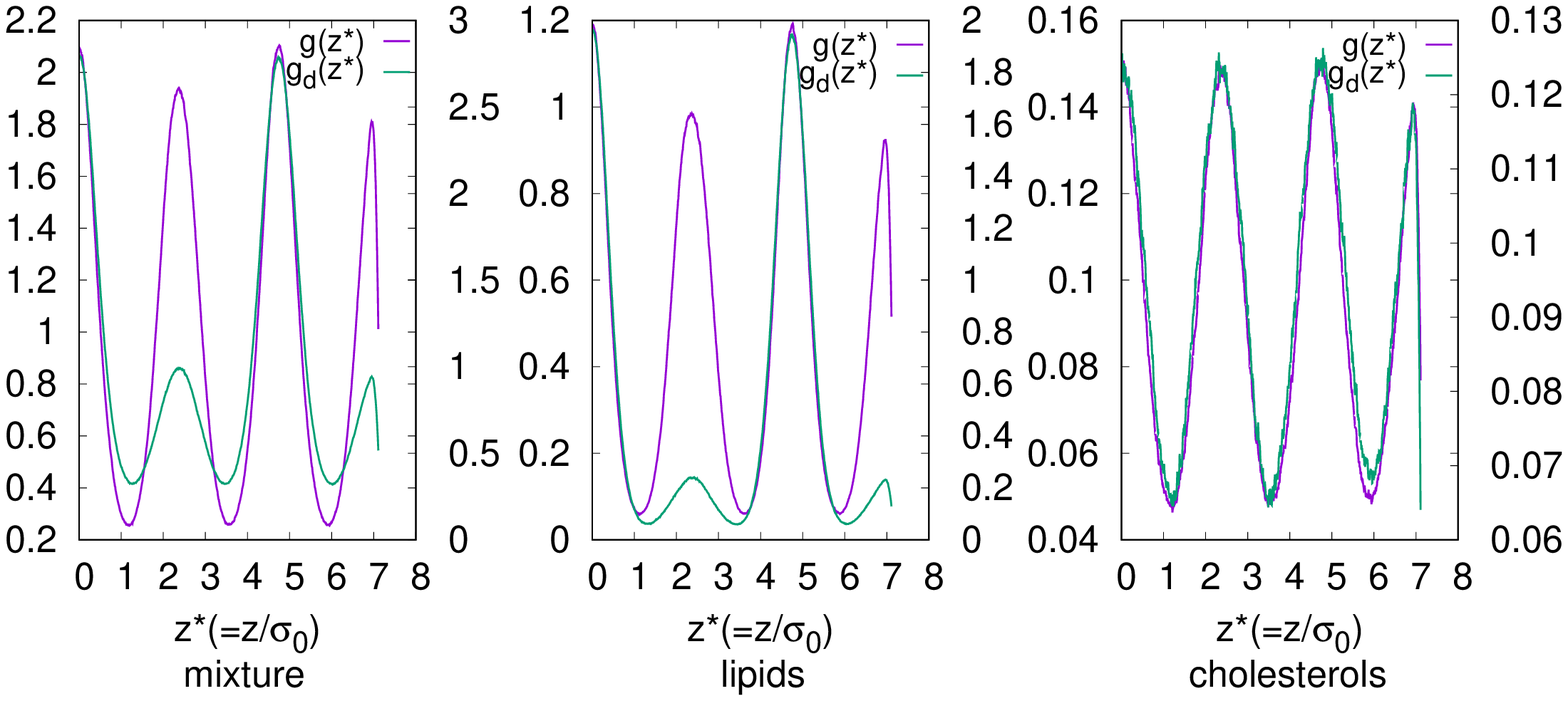}
 \caption{$c=1.5$, $\mu^*_c=0.5$, $\mu^*_l=1.4$.\label{gofz_c1.5_1.4_0.5}}
 \end{subfigure}
 \caption{Plots of $g(z^*)$ and $g_d(z^*)$ for different system with $c=1.5$.\label{gofz_c1.5_0.1and0.5}}
\end{center}
\end{figure}

In these cases with $\mu^*_c=0.1$ and $0.5$, partial bilayer ordering has been found in between the smectic layers. The polar achiral particles with dipole moment $\mu^*_l=1.4$, which is present relatively higher in concentration ($=70\%$) in the system, forms these bilayered smectic domains, whereas, the small smectic domains of chiral polar particles do not form bilayer. The plots of pair correlation functions $g(z)^*$ of molecular center of masses and that of dipolar positions $g_d(z^*)$ along the director axis (figure: \ref{gofz_c1.5_0.1and0.5}), computed in a fashion similar as $g(r^*)$, supports the smectic layer formation. But, if the functions are plotted taking both type of particles into account, then these plots do not show clear existence of bilayered domains (referred to the plots for `mixture' in figure: \ref{gofz_c1.5_1.4_0.1} and \ref{gofz_c1.5_1.4_0.5}). But when they are plotted considering different types of molecules separately, for polar achiral molecules (referred to the plots of `lipids' in the figure: \ref{gofz_c1.5_1.4_0.1} and \ref{gofz_c1.5_1.4_0.5}) alternate peaks of both the functions are of comparable heights, indicating partial presence of bilayer where dipolar parts of polar achiral molecules (the model lipid molecules) of two adjacent smectic layers have been gathered together, whereas, for polar chiral molecules (the model cholesterol molecules) all the peaks of the both functions are of comparable heights (referred to the plots of `cholesterols' in the figure: \ref{gofz_c1.5_1.4_0.1} and \ref{gofz_c1.5_1.4_0.5}). Clearly, this partial bilayer formation is more prominent in the system with $\mu^*_c=0.5$ than $\mu^*_c=0.1$. Snapshots of the configurations obtained in these cases are shown in the figure \ref{qmga_ch1.5_1.4_0.1} and \ref{qmga_ch1.5_1.4_0.5}.

\begin{figure}[h]
\begin{center}
 \includegraphics[width=0.48\textwidth]{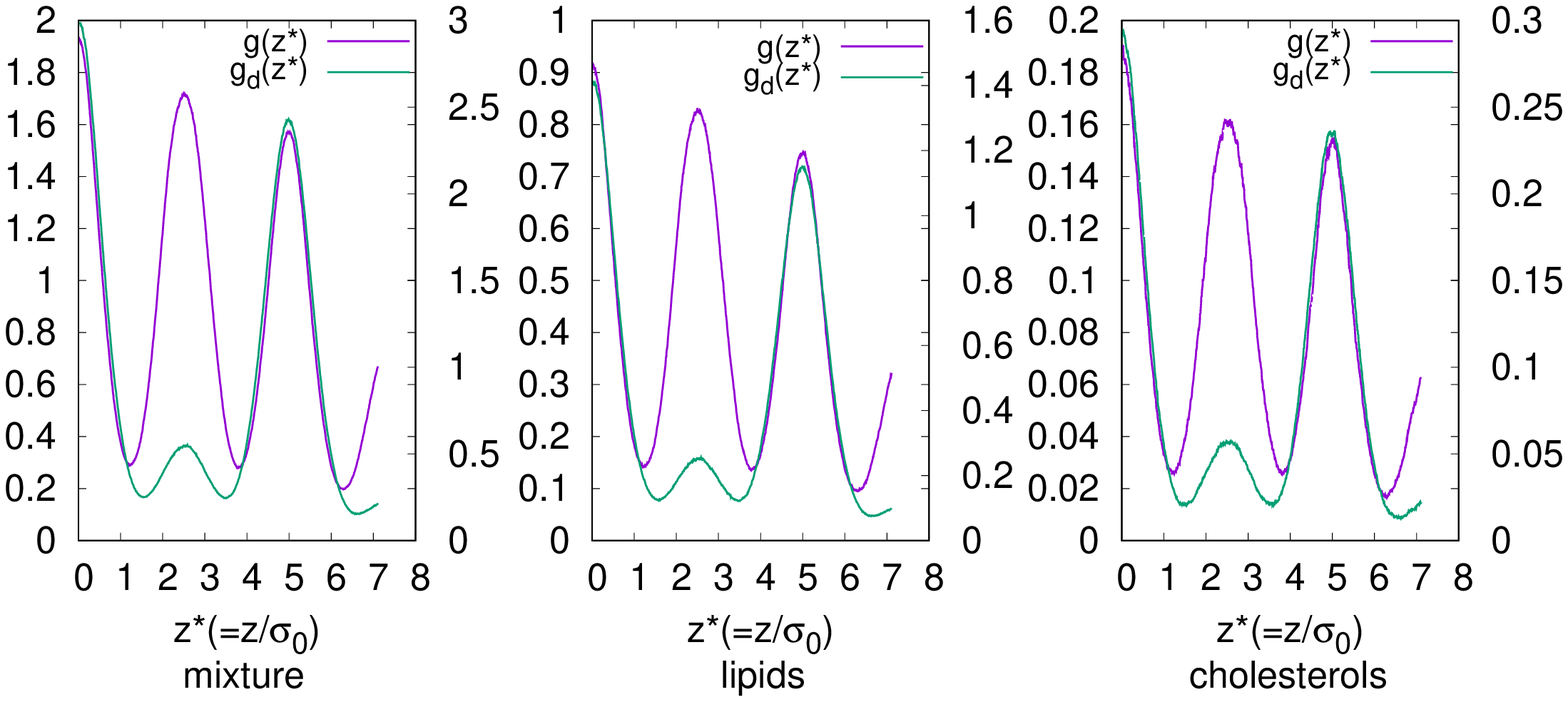}
 \caption{Plots of $g(z^*)$ and $g_d(z^*)$ for the system with $c=1.5$, $\mu^*_c=1.4$, $\mu^*_l=1.4$.\label{gofz_c1.5_1.4_1.4}}
\end{center}
\end{figure}

\begin{figure}[h]
\begin{center}
 \begin{subfigure}{0.48\textwidth}
  \begin{subfigure}{0.44\textwidth}
  \includegraphics[width=\textwidth]{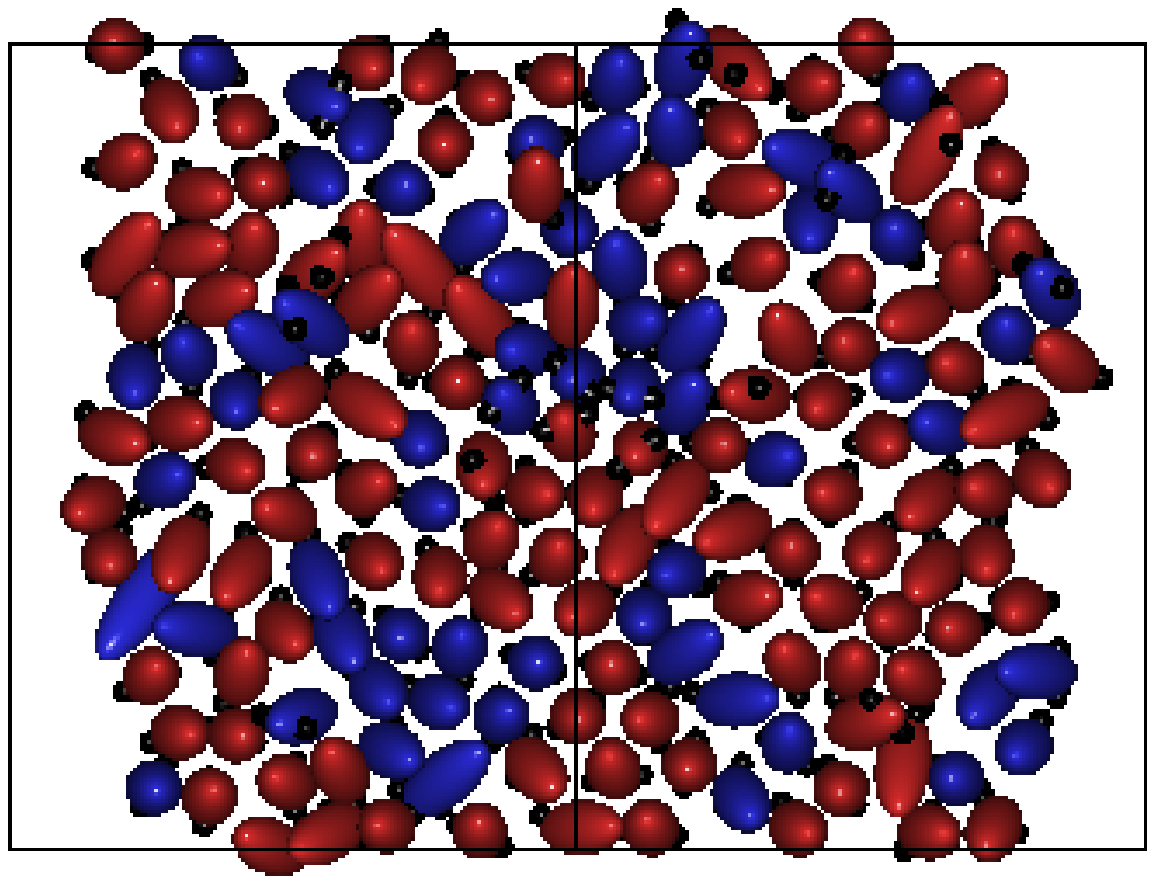}
  \end{subfigure}
  \begin{subfigure}{0.44\textwidth}
  \includegraphics[width=\textwidth]{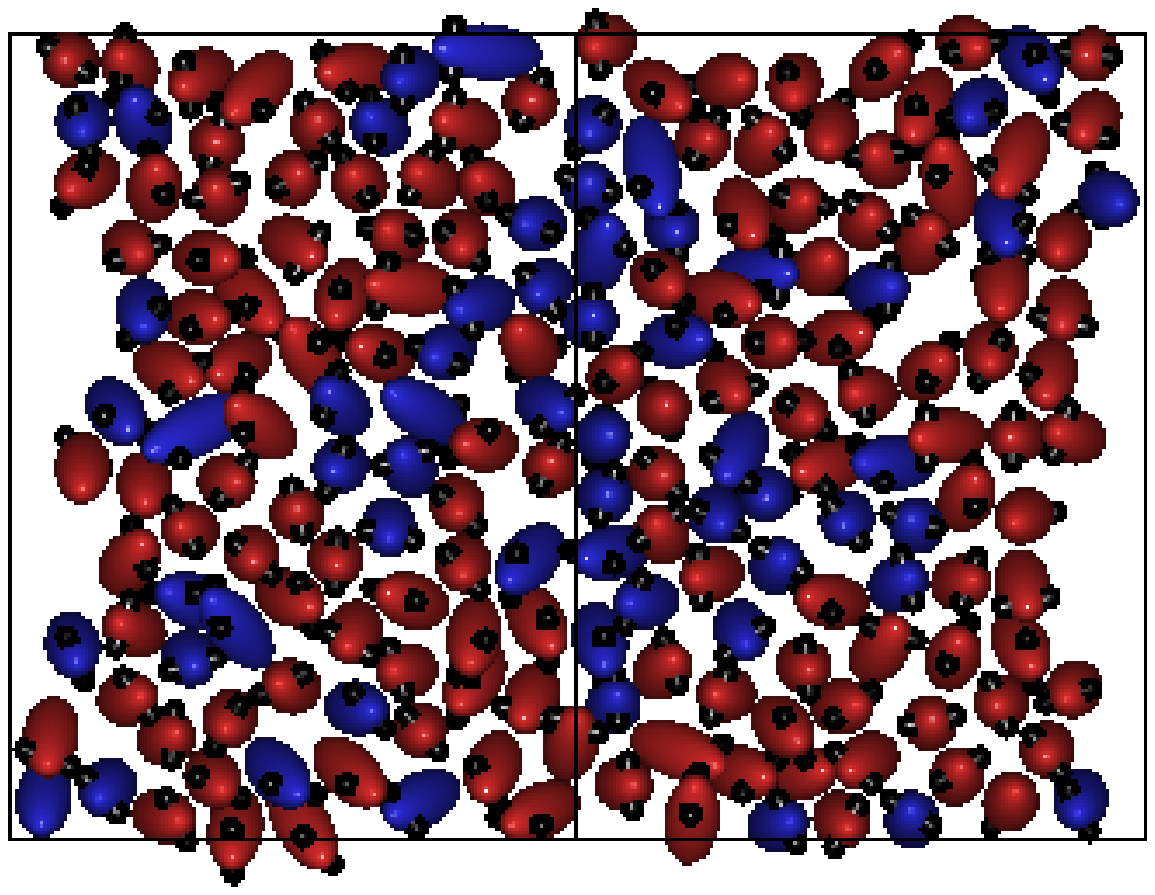}
  \end{subfigure}\\
  \begin{subfigure}{0.44\textwidth}
  \includegraphics[width=\textwidth]{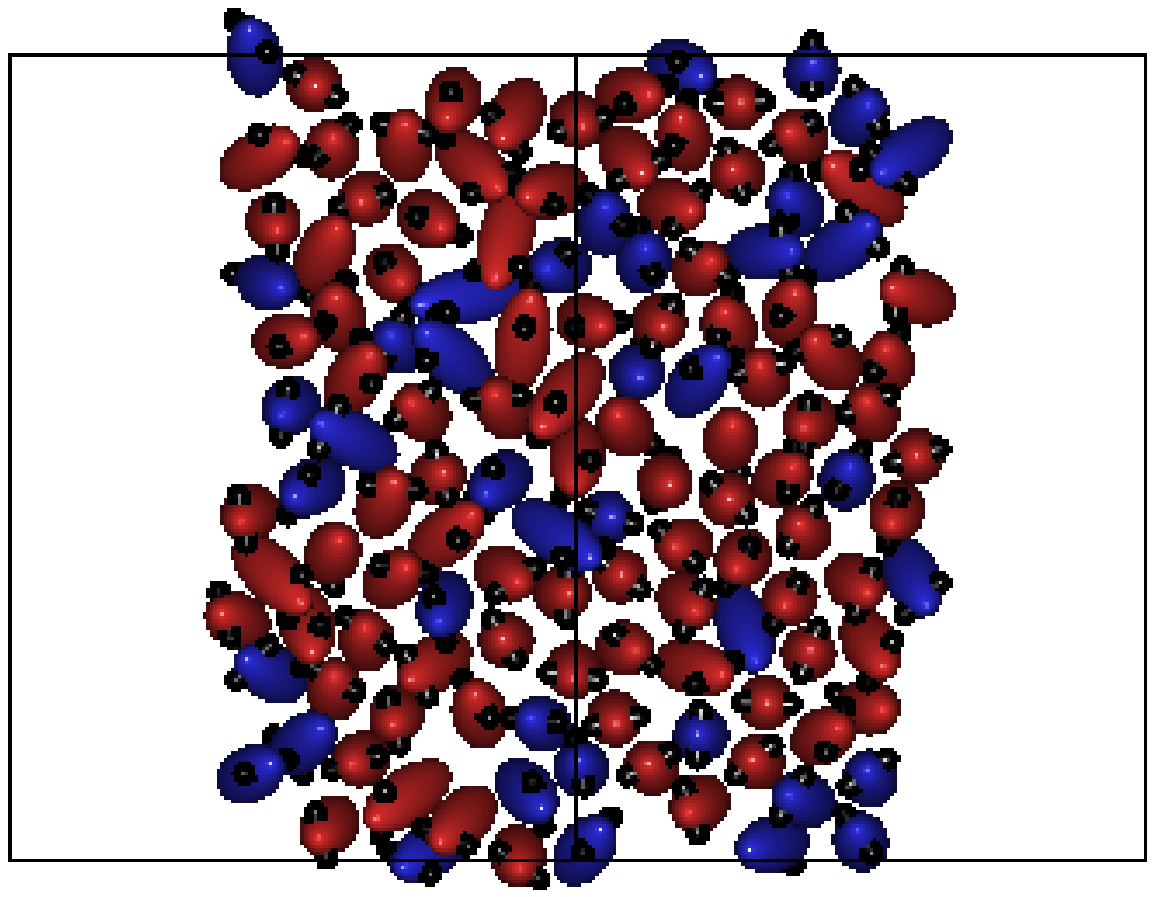}
  \end{subfigure}
  \begin{subfigure}{0.44\textwidth}
  \includegraphics[width=\textwidth]{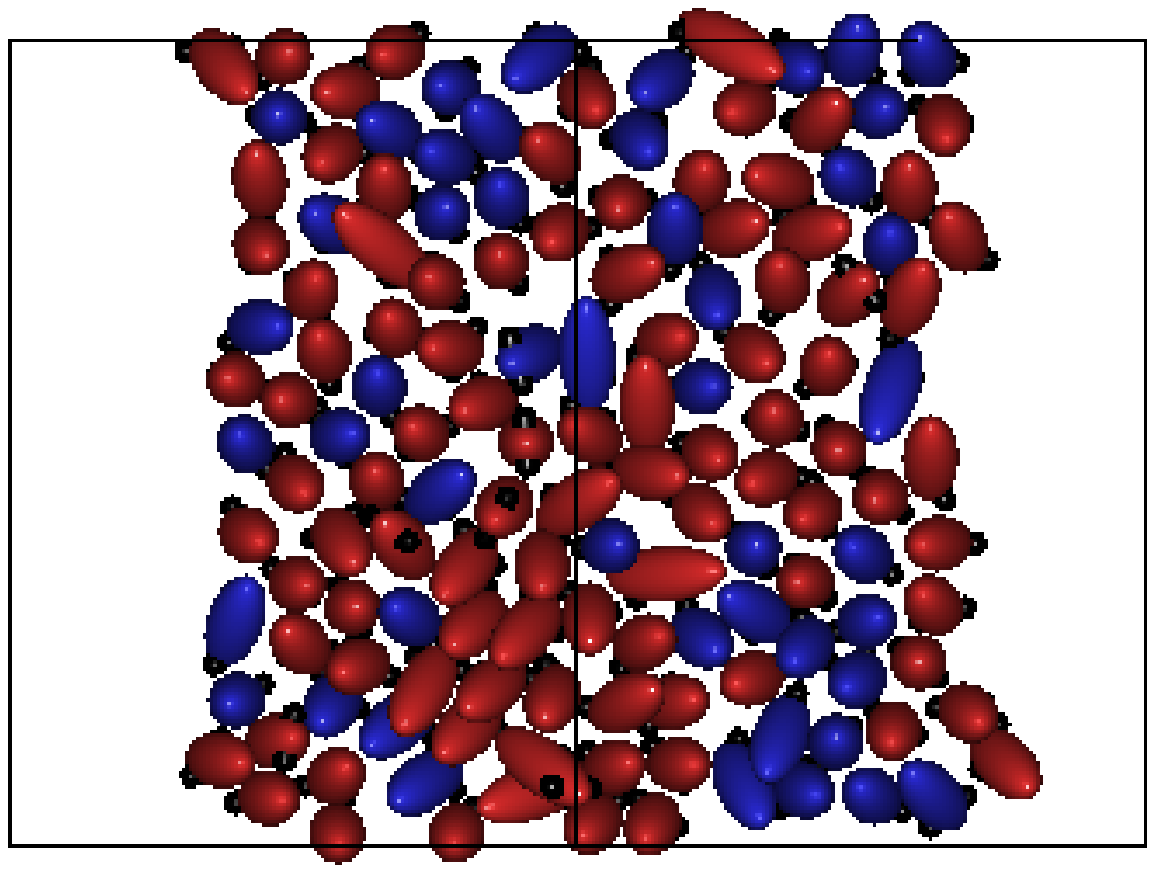}
  \end{subfigure}
 \end{subfigure}
 \caption{Snapshots of the configurations of some of the smectic layers obtained with $c=1.5$, $\mu^*_l=1.4$, (a) $\mu^*_c=1.4$ for $N=864$. View from the top of the layers i.e. along the direction of the director axis has been presented to show the particle positions only. Particles in red represent polar achiral molecules and particles in blue are polar chiral molecules.\label{fig:layers_2}}
\end{center}
\end{figure}

To check the effect of the dipole moment, a simulation run has been performed with the $\mu^*_c$ value same as that of $\mu^*_l$ i.e. both equal to $1.4$. With both the values of $c=1.0$ and $1.5$, in these systems too, nematic phases have been formed (figure: \ref{fig:g2ofr_ch1.0} and \ref{fig:g2ofr_ch1.5}) on decreasing the temperature from respective higher temperature isotropic phases. On further decrease in the temperature, smectic layers have been formed, but, most interestingly, complete bilayered ordering has been found here in the smectic layers, i.e., the dipolar parts of both types of molecules of adjacent smectic layers gathered together to form an uniform bilayered structure. Snapshot of the configurations obtained with such $\mu^*_l$ and $\mu^*_c$ values have been presented in the figure \ref{qmga_ch1.0_1.4_1.4} for a system having $c=1.0$ and in figure \ref{qmga_ch1.5_1.4_1.4} for $c=1.5$. The plots of $g(z^*)$ and $g_d(z^*)$ (figure: \ref{gofz_c1.5_1.4_1.4} with $c=1.5$) for the mixture of both type of molecules and separately for each type of molecules show similar variation. In these cases, the alternate peaks of comparable heights for both the functions have been occurred in same position indicating the presence of completely bilayered smectic layers. Small domains of chiral molecules have also been found to occur in this case. Snapshots of some smectic layers in the configuration obtained in a system with $c=1.5$ have been shown in figure \ref{fig:layers_2}, which shows the presence of small domains riched with polar chiral molecules i.e. the model cholesterol molecules. These domains are also bilayered in this case, as supported by the plots of $g(z^*)$ and $g_d(z^*)$ (figure: \ref{gofz_c1.5_1.4_1.4}). The highest scaled temperature at which this completely bilayered smectic phase has been formed is relatively greater, having value $T^*=1.8$, than that where partially bilayered smectic phases formed with lower values of $\mu^*_c$. For $\mu^*_c=0.1$ and $0.5$, the partial bilayers start forming at a scaled temperature $T^*=1.6$.

In this NVT Molecular Dynamics simulation study, it has been found that, the strength of chiral interaction is an important factor which controls the domain formation in the system of mixture of polar achiral molecules and polar chiral molecules of same sizes. Here, the polar achiral molecules have been used as a coarse-grained model of lipid molecules and the polar chiral molecules as the model of cholesterol molecules. Formation of stable bilayers has been possible controlling the dipole moment of the chiral molecules. Despite the simple coarse-grained modelling of a system of the mixture of lipids and cholesterols, this simulation study successfully generates cholesterol rich small smectic domains in a matrix of lipid multi-bilayer considering molecular-level physical interactions. Thus, this study represents a computational model for the `rafts' that can be found in lipid membranes, the study of which is important not only from the point of its complex structure and phase behaviour but also of biological importance. Further studies focusing on molecular level coarse-grained modelling of certain affinities between different kinds of lipids e.g. phospholipids, sphingolipids etc., cholesterols and proteins are needed to develop complete understanding of the complex phase-behaviour of the plasma membrane. The phase separation between cholesterol and lipid molecules, as has been resulted in the present simulation, may as well be considered as the first step to simulate the organization of rafts. Considering all these points, it is our aim to study multicomponent bilayer, consisted of phospholipids, sphingolipids, cholesterols and proteins present in suitable percentage in appropriate environment to understand the mechanism behind raft formation.

\end{document}